\documentclass[twocolumn,showplace,showpacs,floatfix,prb,footinbib]{revtex4}
\usepackage{amssymb}
\usepackage{graphicx}
\usepackage{dcolumn}
\usepackage{bm}

    \font\tenbifull=cmmib10 
    \font\tenbimed=cmmib7
    \font\tenbismall=cmmib5
\def\bmit{\fam9 }
\mathchardef\bbGamma="7000 \mathchardef\bbDelta="7001
\mathchardef\bbPhi="7002 \mathchardef\bbAlpha="7003
\mathchardef\bbXi="7004 \mathchardef\bbPi="7005
\mathchardef\bbSigma="7006 \mathchardef\bbUpsilon="7007
\mathchardef\bbTheta="7008 \mathchardef\bbPsi="7009
\mathchardef\bbOmega="700A \mathchardef\bbalpha="710B
\mathchardef\bbbeta="710C \mathchardef\bbgamma="710D
\mathchardef\bbdelta="710E \mathchardef\bbepsilon="710F
\mathchardef\bbzeta="7110 \mathchardef\bbeta="7111
\mathchardef\bbtheta="7112 \mathchardef\bbiota="7113
\mathchardef\bbkappa="7114 \mathchardef\bblambda="7115
\mathchardef\bbmu="7116 \mathchardef\bbnu="7117
\mathchardef\bbxi="7118 \mathchardef\bbpi="7119
\mathchardef\bbrho="711A \mathchardef\bbsigma="711B
\mathchardef\bbtau="711C \mathchardef\bbupsilon="711D
\mathchardef\bbphi="711E \mathchardef\bbchi="711F
\mathchardef\bbpsi="7120 \mathchardef\bbomega="7121
\mathchardef\bbvarepsilon="7122 \mathchardef\bbvartheta="7123
\mathchardef\bbvarpi="7124 \mathchardef\bbvarrho="7125
\mathchardef\bbvarsigma="7126 \mathchardef\bbvarphi="7127

\def\boldsigma{\bmit\bbsigma}

\begin{document}
\title{Quantum transport through a deformable molecular transistor}

\author{P. S. Cornaglia}
\author{D. R. Grempel}
\author{H. Ness}
\affiliation{CEA-Saclay, DSM/DRECAM/SPCSI, B\^at. 462, F-91191 Gif
sur Yvette, France}
\begin{abstract}
The linear transport properties of a model molecular transistor
with electron-electron and electron-phonon interactions were
investigated analytically and numerically. The model takes into
account phonon modulation of the electronic energy levels and of
the tunnelling barrier between the molecule and the electrodes.
When both effects are present they lead to asymmetries in the dependence of the
conductance on gate voltage. The Kondo effect is observed in the
presence of electron-phonon interactions. There are important
qualitative differences between the cases of weak and strong
coupling. In the first case the standard Kondo effect driven by
spin fluctuations occurs. In the second case, it is driven by
charge fluctuations. The Fermi-liquid relation between the
spectral density of the molecule and its charge is altered by
electron-phonon interactions. Remarkably, the relation between the
zero-temperature conductance and the charge remains unchanged.
Therefore, there is perfect transmission in all regimes whenever
the average number of electrons in the molecule is an odd integer.
\end{abstract}
\pacs{71.27.+a,75.20.Hr,73.63.-b,85.65.+h}

\maketitle
\section{Introduction}
There has been in recent years a surge of interest in the study
of transport phenomena in nanoscale systems motivated by
their potential as electronic devices. These systems
 differ fundamentally
from conventional conductors in that their electronic spectrum
is discrete and their charging energies can not be ignored.

Charging effects on the transport properties
 have been intensively studied.
Coulomb blockade has been observed in the
conductance of quantum dots~\cite{CB_Qdots} and in single-molecule
devices weakly coupled to the electrodes.
\cite{JPark_2002,Kubatkin_2003} The enhancement of the
low-temperature conductance in the valley between Coulomb blockade
peaks due to the Kondo effect has also been observed both in
quantum dots \cite{Gordon1998} and in single molecules having well
defined spin and charge states.~\cite{JPark_2002,WLiang_2002,Yu2004,Yu2004b}

An important feature of molecules is that they
 generally distort upon the addition or the removal of
electrons. While in conventional systems the energies associated with
atomic motion are much lower than typical electronic energies,
this is not necessarily true in molecular devices.
 It has been recently
shown that the Coulomb charging energies of single molecules can
be considerably reduced by screening due to the electrodes.
\cite{Kubatkin_2003} The former may drop from a few eV, for an
isolated molecule, to a few hundreds of meV becoming then of the
same order of magnitude than vibrational energies. Interesting new
physics emerges when the two energy scales become comparable as
was observed in inelastic electron tunnelling spectra of small
molecules adsorbed on surfaces, \cite{WHoXX} in ${\rm C}_{60}$
molecular-scale transistors,\cite{HPark_2000} and in a suspended quantum dot cavity.\cite{Weig2004}

In this paper we study the linear transport properties of a model
molecular device in which electron-electron and electron-phonon
interactions are present. Previous authors investigated similar
models in the regimes of high temperature or weak electron-phonon
coupling. ~\cite{Mitra_2004,Kuo_2002,Flensberg_2004} We are interested here
in the low-temperature transport properties in regimes for which
the charging energy and the electron-phonon energies can be of the
same order and large compared to the broadening of the electronic
levels. This requires treating both interactions on the same
footing and non-perturbatively.

In a vibrating molecule connected to electrodes the position of
the electronic energy levels with respect to the Fermi level and
the height of the tunnelling barrier between the molecule and the
electrodes are phonon modulated. These effects have important 
consequences for the transport.
In a previous publication~\cite{Cornaglia2004} we gave a short
account of our results for a model in which only the first effect 
was included. Here, we present a detailed analysis of the model,
 including the effects of tunnelling-barrier modulation.

Our results are analytical and numerical. Fermi-liquid theory was
used to derive some general properties of the zero-temperature
conductance and detailed information on its dependence on the
parameters of the model was obtained in the limits of weak and
strong electron-phonon coupling.  Numerical
calculations were also performed in all the relevant parameter
regimes using the Numerical Renormalization Group method  at zero
and at finite temperature.

Our main results are as follows. In all parameter regimes there is
a peak in the zero-temperature conductance as a function of gate
voltage $G(V_g)$ whose hight corresponds
 to perfect transmission through the junction. 
It occurs at the value of $V_g$ for which the average number of
electrons in the device is an odd integer.

The width and shape of the peak depend on the type and strength of
the electron-phonon coupling. When either energy level modulation
(ELM) or tunnelling barrier modulation (TBM) are present (but not
both), $G(V_g)$ is symmetric around its maximum. However, if the
two are simultaneously present, the curve is asymmetric. This
feature is very pronounced in some cases. The two couplings have
opposite effects on the width of $G(V_g)$. ELM leads to peak
narrowing while TBM has the opposite effect.

There are parameter regimes in which resonant transmission is due
to the Kondo effect. When the amplitude of ELM ranges from low to
moderate, the effect is qualitatively similar to that observed in
the absence of the electron-phonon coupling. When ELM is strong,
however, the nature of the ground state of the molecule changes
and the Kondo effect does not result from spin fluctuations  (as
is the case in quantum dots) but from charge fluctuations. There
are important qualitative differences between the two cases.

 At finite temperature, in the Kondo regime, the height
of the conductance peak decreases and is strongly suppressed
 beyond a characteristic temperature, the Kondo temperature.
An increase in TBM by itself leads to an increase of the Kondo temperature. The
effect of increasing ELM is non monotonic: the Kondo temperature increases
for small coupling but it decreases for strong coupling.

Above the Kondo temperature well defined asymmetric Coulomb
blockade peaks are observed if ELM is not too strong. When its
strength increases, the Coulomb peaks become closer to each other
and disappear in the strong coupling regime. The effect of TBM is
to smear these peaks when they exist.

The rest of this paper is organized as follows. In
Section~\ref{sec:model} we describe the model that we use in our
calculations. Section~\ref{sec:analytical} contains our analytical
results, including the derivation of the Fermi-liquid relations
for our model and the analysis of the limiting cases of weak and
strong electron-phonon coupling. Our numerical results are
presented and discussed in Section~\ref{sec:numerical}. Finally,
we state the conclusions of our study in
Section~\ref{sec:conclusions}.

\section{The model}
\label{sec:model} 
\begin{figure}[tbp]
\includegraphics[width=8.5cm,clip=true]{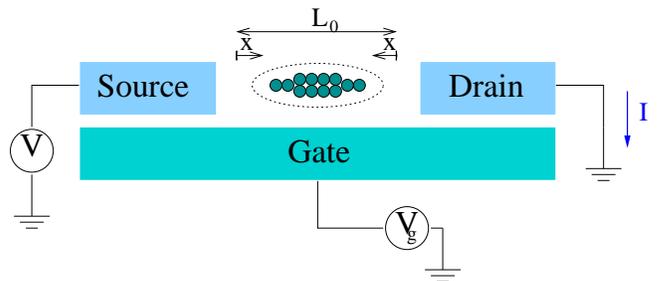}
\caption{A scheme of the model device studied in this paper. A
molecule of average length $L_0$ is connected to source and drain
electrodes. As it vibrates symmetrically, the distance between the
end groups and the electrodes fluctuates thus modulating the
tunnelling barrier between the elements. There is also a
modulation of the position of the molecular energy levels with
respect to the Fermi level not represented in the scheme. }
\label{fig:fig1}
\end{figure}

We model a molecule connected to metallic
electrodes in a range of gate voltages such that only one molecular
orbital effectively participates in the transport. The molecule
has a symmetric vibrational mode of frequency $\omega_0$ coupled
to the electronic coordinates. The energy $\varepsilon_d(x)$ of
the molecular level and the tunnelling matrix elements between the
molecule and left (L) and right (R) electrodes, $V_{\ell}(x) =
V(x)$ ($\ell$ = L,R), depend on a dimensionless vibrational
coordinate $x$. For small distortions these quantities may be
expanded as $\varepsilon_d(x) \approx \varepsilon_d - \lambda\;x$
and $V(x) \approx V\;\left[ 1 + g\;x \right]$ where $\lambda$ and
$g$ are two coupling constants.\cite{Arrachea2003} The first one is an energy scale,
the second one is dimensionless. A scheme of the device is shown
in Fig.~(\ref{fig:fig1})

 The Hamiltonian of the combined
system consists of three terms that describe the isolated
molecule, the isolated electrodes, and the coupling between them,
respectively,
\begin{equation}
H=H_{M} + H_{E} + H_{M-E}\;. \label{eq:hamil}
\end{equation}
The Hamiltonian of the molecule is
\begin{eqnarray}
H_{M}=\varepsilon_d \hat{n}_d + U n_{d \uparrow}n_{d
\downarrow}-\lambda\; \left(\hat{n}_d-1
\right)\left(a^{}+a^{\dagger}\right)+\omega_0a^{\dagger}a^{}\;,
\label{eq:molecule}
\end{eqnarray}
where $\hat{n}_d = \sum_{\sigma} d^{\dagger}_\sigma d^{}_\sigma$ is the
charge operator and $U$ is the Coulomb repulsion between two
electrons on the same molecular orbital. In appropriate units we
write the elongation of the molecule as $x = a + a^{\dagger}$,
where $a$ and $a^\dagger$ are the phonon operators.

The Hamiltonian of the leads is
\begin{eqnarray}
H_{E}&=&\sum_{{\bf k},\sigma,\ell} \varepsilon_{\ell}({\bf k})\;
c^{\dagger}_{{\bf k} \sigma \ell} c^{}_{{\bf k} \sigma
\ell}\;\left(\ell={\rm L,R}\right)\;, \label{eq:leads}
\end{eqnarray}
where we consider for simplicity the case of identical electrodes
with dispersion $\varepsilon_{\textsc{r}}({\bf k}) =
\varepsilon_{\textsc{l}}({\bf k}) =\varepsilon({\bf k})$.
We assume for convenience that the conduction band is symmetric
with respect to the chemical potential and we take the latter
as the origin of energies.

The third term in Eq.~(\ref{eq:hamil}) is the tunnelling
Hamiltonian between the molecule and the electrodes,
\begin{eqnarray}
H_{M-E}&=&[1 + g\;(a + a^{\dagger})]V \sum_{{\bf
k},\sigma,\ell}\;\left(d^{\dagger}_{\sigma} \;c^{}_{{\bf
k}\sigma\ell} + {\textsc{h.c}}
\right) \\ \nonumber &\equiv& \left[ 1 + g\;(a + a^{\dagger})
\right] V \hat{v}\;, \label{eq:tunnelling}
\end{eqnarray}
where we assumed that the tunnelling matrix element $V$ is
$\bf{k}$-independent and we defined the operator $\hat{v}$ for
notational convenience.

Note that if either $g=0$ or $\lambda=0$ and  $\epsilon_d = -U/2$,
the Hamiltonian of the model is symmetric under an
 electron-hole transformation  (plus the inversion $x \to -x$
 if $\lambda \ne 0$). At the symmetric point, the average number of electrons of the molecule is $n_d\equiv\langle\hat{n}_d\rangle=1$.
If both $g$ and $\lambda$ are non vanishing, this symmetry is
lost. We shall see below that this has important consequences.

An expression for the conductance $G$ of the molecular junction at
zero bias can be derived using standard
methods.~\cite{Meir1992,Jauho1994} We find
\begin{equation}
\label{eq:cond-def} G=\left.{dI\over dV}\right|_{V=0}={2e^2\over
h}\ \pi \Delta
 \int_{-\infty}^{\infty} d\omega
 \left(-{\partial f(\omega)\over \partial \omega}\right)
 \widetilde{\rho}_{dd}({\omega})\;,
\end{equation}
where $f(\omega)$ is the Fermi function, $\Delta = 2 \pi \rho_0
V^2$ and $\rho_0$ is the electronic density of states of the
electrodes evaluated at the Fermi level. The spectral density
$\widetilde{\rho}_{dd}(\omega)=-\pi^{-1}{\rm
Im}\widetilde{G}_{dd}(\omega)$,  where we defined a  modified
$d$-electron Green function $\widetilde{G}_{dd}(\omega)$ as
\begin{eqnarray}
\label{eq:G-tilde} \widetilde{G}_{dd}(\omega)&=& -i \int_0^\infty dt
e^{i \omega t}\Big{\langle} \left[\left(1 + g\;x(t)\right)
d_\sigma(t),\right. \nonumber\\
&&\left. \left(1 + g\;x\right) d^{\dagger}_\sigma \right]_{+}
\Big{\rangle}\;.
\end{eqnarray}

\section{Analytical results}
\label{sec:analytical}

\subsection{Exact results at zero temperature}
\label{sec:Fermi}

The ground state of Hamiltonian~(\ref{eq:hamil}) is expected to be
a Fermi liquid. Exact relationships can then be established
between the zero-temperature conductance through the junction, the
spectral density of the molecule at the Fermi level and its
charge. To derive these relations the structure of the expansion
of the electronic Green functions in powers of the
electron-electron and electron-phonon interactions coupling
constants must be examined.

We start by noting that when the molecule is coupled to
the contacts it will in general deform. It is convenient to carry
out the perturbation expansion around the deformed ground state.
We thus apply a canonical transformation that shifts the phonon
operators, $a \to b + \alpha$, and write
\begin{equation}
H(a,a^{\dagger}) = H(b + \alpha,b^{\dagger} + \alpha) \equiv
H_\alpha^\prime (b,b^{\dagger})\;, \label{eq:H-shifted}
\end{equation}
which defines the transformed Hamiltonian $H_\alpha^\prime
(b,b^{\dagger})$.

Since the ground state energy is invariant under this transformation,
\begin{equation}
\label{eq:canonical} E_0 \equiv \langle H \rangle = \langle
H_\alpha^\prime \rangle\;,
\end{equation}
we may take the derivative with respect to $\alpha$ of this equation
 to obtain,
\begin{eqnarray}
\omega_0\;\langle b +
b^{\dagger}\rangle_\alpha + 2\alpha \omega_0 - 2\lambda \langle
\hat{n}_d -1 \rangle_{\alpha} + 2 g V \langle \hat{v} \rangle_{\alpha}
\equiv 0\;,  \label{eq:H-invariance}
\end{eqnarray}
where $\langle \cdots \rangle_\alpha$ denotes an expectation value
 with respect to the {\it exact} ground state of the transformed Hamiltonian.
The deformation $\langle x \rangle$ of the molecule is found by
imposing that $\langle b + b^{\dagger}\rangle_{\alpha} =0 $ on the
transformed ground state:
\begin{equation}
\alpha = \frac{\langle x \rangle}{2}=\frac{\lambda}{\omega_0}
\langle \hat{n}_d -1 \rangle_{\alpha} - \frac{g}{\omega_0} V \langle
\hat{v} \rangle_{\alpha}\;.
 \label{eq:elongation}
\end{equation}
The transformed Hamiltonian can be expressed as a sum of one-body
 and many-body terms as
\begin{eqnarray}
\label{eq:H0+H1}
H^\prime &=& H^\prime_0 + H^\prime_1\;,\\
\nonumber \\ \label{eq:H0} H^\prime_0 &=& H_E + \omega_0\;
b^{\dagger}b + \widetilde{\varepsilon}_d\;\hat{n}_d +  \widetilde{V}\;\hat{v}\;, \\
\nonumber \\ \label{eqq:H1} H^\prime_1&=& \left(b + b^{\dagger}\right)
\left[ g V \left(\hat{v}-\langle \hat{v} \rangle_{\alpha}\right) -
\lambda \left(\hat{n}_d-\langle \hat{n}_d \rangle_{\alpha}\right)\right]\\
\nonumber &+& U n_{d \uparrow} n_{d \downarrow} \;,
\end{eqnarray}
where we dropped a constant energy shift and we defined the
renormalized parameters
\begin{eqnarray}
\label{eq:renorpar} \widetilde{\varepsilon}_d = \varepsilon_d - 2
\lambda \alpha\;,\;\textrm{and}\;\;\widetilde{V} =  V \left( 1 + 2
g \alpha \right)\;.
\end{eqnarray}

The molecule is only coupled to a local symmetric
linear combination of the states of the leads,
\begin{equation}
\left|\Psi_c\rangle\right. =  {\cal N}^{-1/2} \sum_{\bf k}
\frac{\left|\Psi_{\bf{k},\rm{R}}\rangle\right. +
\left|\Psi_{\bf{k},\rm{L}}\rangle\right.}{\sqrt{2}}\;,
\label{eq:s-orbital}
\end{equation}
where ${\cal N}$ is a normalization factor.
The remaining conduction-electron states can be
integrated out reducing the problem to an effective
interacting two-site model with a damping term that reflects
the coupling of the subsystem to the electrodes.
Physical quantities may be obtained from the
 $2\times 2$ Green-function matrix
\begin{eqnarray}
{\bf G}(\omega)=\left(
\begin{array}{cc} G_{cc}(\omega) &  G_{cd}(\omega)\\ \\
G_{dc}(\omega) & G_{dd}(\omega)
\end{array}
\right)\;. \label{eq:green-def}
\end{eqnarray}
The Green function associated to the one-body Hamiltonian $H^\prime_0$ is
\begin{eqnarray}
{\bf G}_0(\omega)=\left(
\begin{array}{cc} 1/G^0_{cc}(\omega) & -\sqrt{2}\;\widetilde{V} \\
\\
-\sqrt{2}\;\widetilde{V} & \omega - \widetilde{{\epsilon}}_d
\end{array}
\right)^{-1}\;, \label{eq:green0}
\end{eqnarray}
where, in the wideband limit, $G^0_{cc}(\omega)= \sum_{\bf{k}}
1/(\omega - \varepsilon_{\bf k}) = -i \pi \rho_0$ and $\rho_0$ is
the bulk density of states of the electrodes.
 Note that ${\bf G}_0(\omega)$ is
{\it not} the non-interacting Green function and it includes
effects of the interactions to all orders through the renormalized
parameters
 (\ref{eq:renorpar}).

The remaining effects of the interaction $H^\prime_1$ are
embodied in a self-energy matrix
defined through the Dyson equation
\begin{equation}
{\bf G}^{-1}(\omega)= {\bf G}_0^{-1}(\omega) - {\bf
\Sigma}(\omega)\;. \label{eq:Dyson}
\end{equation}
In a Fermi-liquid ground state all the elements of the self-energy matrix
 are purely real at the Fermi level. We checked that this is the case for our model
 by computing a few low-order diagrams in the expansion of ${\bf{\Sigma}}$ in powers
 of $H^\prime_1$.  Luttinger's
 theorem~\cite{Abrikosov-book,Hewson-book}can then be generalized
 to the present situation in order to establish some useful Fermi-liquid
relationships.

We found a simple relationship between $\rho_{cc}(0)$, the
$c$-component of the spectral density at the Fermi level, and the
electronic population of the molecule in the physically relevant
wideband limit.
 It reads,
\begin{equation}
\rho_{cc}(0) = \rho_0 \cos^2\left(\frac{\pi}{2}
 n_d\right)\;.
\label{eq:Luttinger2}
\end{equation}
Equation (\ref{eq:Luttinger2}) is an exact result valid for
  all values of the parameters of the Hamiltonian (\ref{eq:hamil}).
For $g=0$ but keeping $\lambda$ arbitrary we found an equivalent
expression for $\rho_{dd}(0)$, the projection of the spectral
density on the molecular orbital. In the wideband limit this is
\begin{equation}
\left.\rho_{dd}(0)\right|_{g=0} = \frac{1}{\pi \Delta}
 \sin^2\left(\frac{\pi}{2} n_d\right)\;,
\label{eq:Luttinger3}
\end{equation}
The content of Eq.~(\ref{eq:Luttinger3}) is that, for $g=0$,
 the $d$-spectral density of the interacting system at
 the Fermi level is pinned at the value it takes for a
non-interacting system with the same electron occupancy. This is a
well known result for model (\ref{eq:hamil})
 in the absence of electron-phonon coupling.~\cite{Hewson-book} We see that
 it remains valid for all values of $\lambda$ and $g=0$.

However, Luttinger's theorem does {\it not} lead to a similar result
 for $\rho_{dd}$ in the general $g\ne 0$ case and
 the elements of the self-energy matrix appear explicitly in the
corresponding expression.

There is some simplification in the particular case $\lambda =0$,
 $\epsilon_d = -U/2$. Then, the Hamiltonian (\ref{eq:hamil}) is electron
 hole symmetric  [c.f. Section \ref{sec:model}], $n_d = 1$,
and $\Sigma_{cc}(0)$ and $\Sigma_{dd}(0)$ vanish. In this case we find
\begin{equation}
\label{eq:rhodd0}
 \left.\rho_{dd}(0)\right|_{\lambda=0,n_d=1} = \frac{1}{\pi
\widetilde{\Delta}}
\end{equation}
where
\begin{eqnarray}
\widetilde{\Delta} = \pi \rho_0 \left[\sqrt{2}\;\widetilde{V} +
\Sigma_{cf}(0)\right]^2\;.
\label{eq:Deltatilde}
\end{eqnarray}
It can be easily shown that $\Sigma_{cf}$ is of order $g^2$ and higher. It then
follows from
Eqs.~(\ref{eq:elongation}), (\ref{eq:renorpar}) and~(\ref{eq:Deltatilde}) that
$\widetilde{\Delta} = \Delta \left(1 + {\cal{O}}(g^2) + \cdots\right)$.

Interestingly, the non-universality of $\rho_{dd}(0)$ (of which
Eq.~(\ref{eq:rhodd0}) is only an example) does not extend to the
$T=0$ conductance. The latter is obtained setting $T=0$ in
Eq.~(\ref{eq:cond-def}).
\begin{equation}
\frac{G}{G_0} = \pi\Delta
\;\widetilde{\rho}_{dd}(0)\;, \label{eq:GTzero}
\end{equation}
where $G_0 =2e^2/h$ is the quantum of conductance.

An expression for the modified Green function $\widetilde{G}_{dd}(\omega)$
can be obtained by
writing down the equation of motion for the conduction-electron
Green function $G_{cc}(\omega)$ using  the definition
 (\ref{eq:G-tilde}) and the original
Hamiltonian~({\ref{eq:hamil}}). We find
\begin{equation}
G_{cc}(\omega) = G^0_{cc}(\omega) + 2 V^2 G^0_{cc}(\omega)
\widetilde{G}_{dd}(\omega)G^0_{cc}(\omega)\;.
\label{eq:eq-of-motion}
\end{equation}
Setting $\omega =0$ and taking the imaginary part of the above
equation this becomes
\begin{equation}
\rho_{cc}(0)=\rho_0 \left [1 - \pi \Delta \widetilde{\rho}_{dd}(0)
\right]\;. \label{eq:rhocctilde0}
\end{equation}
Comparing  Eq.~(\ref{eq:rhocctilde0}) with the Fermi-liquid
relationship (\ref{eq:Luttinger2}) we find
\begin{equation}
\pi \Delta \widetilde{\rho}_{dd}(0) \equiv \frac{G}{G_0} =
\sin^2\left(\frac{\pi}{2}
 n _d\right)\;.
\label{eq:rhoddtilde0}
\end{equation}
The zero-temperature conductance thus depends only on the
 occupation of the molecular orbital just as
 in the absence of tunnelling
barrier modulation and it reaches its maximum value $G_0$ when $n_d=1$.  
A related result was obtained in Ref. \onlinecite{Mitra_2004} in perturbation
 theory in $\lambda$ for $U=0$ and $g=0$.

\subsection{Analysis of limiting cases}

Many features of the solution of our problem  can be found from an
analysis of the limiting cases of small and large electron-phonon
coupling.

We start from the Hamiltonian of the isolated molecule given in
Eq.~(\ref{eq:molecule}). Its eigenvalues and eigenfunctions can be
written down explicitly:\cite{Cornaglia2004}
\begin{eqnarray} \label{eq:enes}
\begin{array}{ll}
\left |0,m\right> = \widetilde{U}^{+}\left |0\right>_e
\left|m\right>,
& E_{0,m}=-{\lambda^2\over \omega_0} + m\omega_0,\\ \\
\left |\sigma,m\right> = \left
|\sigma\right>_e\left|m\right>,
 & E_{\sigma,m}=\varepsilon_d
+ m\omega_0,\\ \\
\left |2,m\right> = \widetilde{U}^{-}\left |\uparrow
\downarrow\right>_e\left|m\right>, & E_{2,m}=-{\lambda^2\over
\omega_0} + 2\varepsilon_d + U + m\omega_0,
\end{array}
\end{eqnarray}
where the subscript $e$ denotes electronic states,
$\left|m\right>$ is $m$-th excited state of the harmonic
oscillator and
\begin{equation}
\widetilde{U}^{\pm} =\exp{\left[\pm{\frac{\lambda}{\omega_0}}
\left(a^{}-a^{\dagger}\right)\right]}\;.
 \label{eq:Upm}
\end{equation}

At $\varepsilon_d = -U/2$ the states with zero and two electrons
are degenerate. We take this as a reference point and write
\begin{equation}
\varepsilon_d = - \frac{U}{2} +  V_d  \;. \label{eq:Vgate}
\end{equation}

The number of electrons in the ground state of the molecule is
either odd or even depending on whether $2\lambda^2/\omega_0$ is
smaller or larger than $U$. It is convenient to define an
 effective interaction parameter,
\begin{equation}
U_{\textrm{eff}} = U - \frac{2\lambda^2}{\omega_0 }
\label{eq:Ueff}
\end{equation}
It will be seen below that the physics in the cases $U_{\textrm{eff}}>0$ and
$U_{\textrm{eff}} < 0$ is quite different.

\subsubsection{$U_{\textrm{eff}} > 0$}
\label{sec:weak-coupling} In this case the ground-state of the
isolated molecule for $|V_d| < U/2$ is the spin-doublet
$\left|\sigma,0\right>$. There is a charge excitation gap given by
$\sim U_{\rm eff}= U - 2 \lambda^2/\omega_0$ for $V_d=0$ and a
phonon gap $\omega_0$. The low-energy excitations are spin
fluctuations.

An effective Hamiltonian for excitations involving the spin doublet
can be derived by perturbatively
 projecting out the empty and doubly occupied
 states from the space of available states by means of a Schrieffer-Wolff transformation.
To lowest order in $V/U$, the low-energy effective
 Hamiltonian for spin fluctuations $H_{\rm eff}$ is
\begin{equation}
H_{\rm{eff}} = \omega_0 a^{\dagger} a + H_E + J {\bf{S}}
\cdot {\bf{s}}_c\;,
 \label{eq:HK-g0}
\end{equation}
where $\bf{S}$ is the spin operator for  the doublet and
${\bf{s}}_c =
1/2\;\sum_{s,s^\prime}\;c^{\dagger}_{s}\;{\boldsigma}_
{s,s^\prime}\;c{}_{s^\prime}$ is the spin operator of an electron
on the state (\ref{eq:s-orbital}) coupled to the molecule.
 The coupling constant $J$ is given by
\begin{equation}
J = \frac{J_0}{2}
\;\sum_{m=0}^\infty~\sum_{\mu =\pm} {|\langle 0 |[1 +
 g (a + a^{\dagger})]
\widetilde{U}^{\mu} | m \rangle |^2\over 1 - 2 \lambda^2/\omega_0 U
 + 2 m \omega_0/U},
\label{eq:J-spin-doublet}
\end{equation}
where $J_0 =8\Delta/(\pi U \rho_0)$ and we have taken $V_d=0$ for simplicity.

A simple analytical expression can be derived in the limit of
small electron-phonon coupling, $g, \lambda/U \ll 1$. We find \footnote{For a related result in the case $g=0$ see Ref. \onlinecite{Stephan1997}. }
\begin{equation}
J = J_0 \left[ 1 + {\left(2\lambda/U\right)^2 + g^2 \over 1 +
2\omega_0/U} + \cdots\right]\;\;.
 \label{eq:Jsmall-lambda-g}
\end{equation}
Therefore, $J$ increases with the electron-phonon coupling
in the weak coupling limit.

Equation~(\ref{eq:HK-g0}) is the well known Kondo model Hamiltonian
.~\cite{Hewson-book} The dependence of the conductance
on temperature, gate-voltage and magnetic field is well
understood in the absence of electron phonon coupling.
We summarize below the main features for $g=\lambda=0$.

At $T=0$, there is a narrow resonance of width $T_K \sim D
\exp\left(-1/(J\rho_0)\right)$ in the $d$-electron spectral
density at the Fermi level. This resonance
 provides a channel for conduction and, at the symmetric point, $n_d = 1$,
 $G = G_0$ in agreement with Luttinger's theorem.

For $V_d \ne 0$, the gap between the ground state doublet and the
empty state ($V_d > 0$), or the doubly occupied state ($V_d < 0$)
is reduced. The charge then deviates from $n_d = 1$. The Kondo
resonance shifts with respect to the Fermi level
  and the $\rho_{dd}(0)$ decreases.
For  $|V_d| \gtrsim U/2$ the Kondo resonance disappears.
 There is thus a peak in $G(V_g)$ whose width
$\sim U$ is large. An applied magnetic field $B$ also
destroys the Kondo resonance by breaking the symmetry between the
spin up and spin down ground states. This happens for $B \sim T_K$
which is very small in the Kondo limit. Therefore, the peak in
$G(B)$ is narrow.
At finite temperature the conductance at the symmetric point
 decreases becoming very small for $T \gtrsim T_K$. In
this temperature range $G$ exhibits Coulomb blockade peaks at
gate voltages $V_d \sim \pm U/2$.

These features are present throughout the weak coupling regime with some
 minor modifications. It follows from equation~(\ref{eq:Jsmall-lambda-g})
that $T_K$ increases with increasing $\lambda$ and $g$.
Therefore, the peak in $G(B)$ becomes wider and the temperature variation of
 the conductance at resonance becomes slower.
If both $g$ and $\lambda \ne 0$ the zero-temperature peak in
$G(V_d)$ shifts from $V_d = 0$ to $V_d = V^\star_d$ such that $n_d
=1$ and acquires an asymmetric shape. These features are  related
to the loss of particle-hole symmetry referred to above. The
conductance is higher to the left of the maximum than to its right
 because for negative values of $V_d - V^\star_d$
 the doubly occupied state that
 has a stronger coupling to the electrodes is favored.
The width of the peak and  the separation between the
finite-temperature Coulomb blockade peaks, that are also
asymmetric, decrease with increasing $\lambda$ mirroring the
evolution of $U_{\textrm{eff}}$.

\subsubsection{$U_{\textrm{eff}} < 0$}
\label{sec:strong-coupling} In this regime the ground-state for
$V_d = 0$ is a charge doublet formed by the states
$\left|2,0\right>$ and $\left|0,0\right>$. The low-energy
excitations of the system are thus charge fluctuations and there
is a large gap $\sim 2\lambda^2/\omega_0 -U$ for spin excitations.

We discuss the case $g=0$ first. A low-energy effective
Hamiltonian can be found as before applying a Schrieffer-Wolff
transformation, but now we project out  the singly occupied
states.  We
 introduce pseudo-spin operators $\tau^z_d$ and $\tau^z_c$
with eigenvalues $\pm 1/2$, corresponding to the eigenvalues $Q=2$ and
 $Q=0$ of the charge of the molecule and of the orbital $c$, respectively.
We also define raising and lowering
 operators
 , $\tau^+_d =\left( \tau^-_d\right)^\dagger =
 d^\dagger_{\uparrow} d^\dagger_{\downarrow}$ and similar ones for the leads.
In terms of these, the effective Hamiltonian for $V_d = g = 0$ is:
\begin{equation}
H = H_ E +
J_{\parallel}\;\tau^z_d\;\tau^z_c + \frac{J_{\perp}}{2}
 \left( \tau^+_d\;\tau^-_c +\tau^-_d\;\tau^+_c  \right)\;,
\label{eq:H-AKM}
\end{equation}
with the couplings
\begin{eqnarray}
\label{eq:Jpara} J_{\parallel} & = & J_0 \sum_{m=0}^\infty
{\langle 0 | \widetilde{U}^{+}| m \rangle \langle m |
\widetilde{U}^{-}| 0\rangle \over
 {2\lambda^2/\omega_0 U} -1 + {2 m \omega_0/U}}\;,\\
\nonumber \\
\label{eq:Jperp} J_{\perp} & = & J_0 \sum_{m=0}^\infty {\langle 0
| \widetilde{U}^{+}| m \rangle \langle m | \widetilde{U}^{+}|
0\rangle\over
 {2\lambda^2/\omega_0 U} -1 + {2 m \omega_0/U}}\;.
\end{eqnarray}
Equations~(\ref{eq:H-AKM})-(\ref{eq:Jperp}) define the Hamiltonian
of  an anisotropic Kondo model (AKM).\cite{Costi1996,Costi1998}
The couplings (\ref{eq:Jpara}) and (\ref{eq:Jperp}) can be easily
estimated in the limit $\lambda/\omega_0 \gg 1$ by noting that
\begin{eqnarray}
\label{eq:U0m} \langle 0 | \widetilde{U}^{+}| m \rangle \langle m
| \widetilde{U}^{-}| 0\rangle &=&
e^{-\left(\lambda/\omega_0\right)^2}\;
\frac{\left(\lambda/\omega_0\right)^{2m}}{m!}  \\ \nonumber
&=&(-1)^m \langle 0 | \widetilde{U}^{+}| m \rangle \langle m |
\widetilde{U}^{+}| 0\rangle \;.
\end{eqnarray}
Considered as  a function of $m$,
 the last expression on the right hand side of the upper line of
Eq.~(\ref{eq:U0m})
is strongly peaked at $m^\star =(\lambda/\omega_0)^2 \gg 1$, while
the denominators in Eqs~(\ref{eq:Jpara}) and (\ref{eq:Jperp}) are
slowly varying functions of $m$. In the strong coupling limit
these can be approximated by their value at $m^\star$ and taken
out of the sum. This results in
\begin{equation}
\label{eq:JAKMlimit} J_{\parallel} \sim  \frac{4 J_0 \omega_0
U}{\lambda^2}\;,\;\;\;J_{\perp} \sim J_{\parallel}  e^{-2
\left(\lambda/\omega_0\right)^2}\;.
\end{equation}

This strong anisotropy originates in the fact that the phonon
ground states corresponding to the two charge states $Q_d =2$ and
$Q_d=0$ are very different. In the strong-coupling limit, the
anisotropy ratio $J_{\perp}/J_{\parallel}$ is precisely the
overlap $\langle 2,0 |0,0 \rangle$ between them [c.f.
Eq.~(\ref{eq:enes})].  The Kondo temperature of the AKM
is~\cite{Costi1996, Costi1998}
\begin{equation}
T_{{\textsc{akm}}} \sim \left({J_\perp \over J_\parallel}\right)
^{1\over J_\parallel \rho_0} \sim D \exp\left[- {\pi \omega_0
\over \Delta} \left({\lambda \over \omega_0}\right)^4\right]\;,
\label{eq:TAKM}
\end{equation}
where the last expression is its asymptotic form for large
$\lambda/\omega_0$.

In contrast with the weak-coupling situation, $T_{\textsc{akm}}$
decreases sharply when $\lambda$ increases. Application of a small
potential $|V_d| \gtrsim T_{\textsc{akm}}$ splits the charge
doublet and destroys the Kondo resonance. Hence, the peak in
$G(V_d)$ is very narrow in this limit. Conversely, the doublet is
insensitive to a magnetic field and the width  $\sim
2\lambda^2/\omega_0 -U$ of the peak in $G(B)$ is now large.
Finally, no Coulomb blockade peaks are to be seen upon application
of a gate voltage because it leads to an {\it increase} of the
energy difference between a ground state with zero or two
electrons and an excited state with one electron irrespective of
its sign. These features are opposite of those that characterize
the properties of $G$ in the weak coupling regime.

We discuss now the case $g\ne 0$. The modulation of the tunnelling
amplitude has two effects. The first one is to renormalize the
coupling constants,
\begin{eqnarray}
\nonumber J_{\parallel}\;&\rightarrow&\;J^\prime_{\parallel} =
J_{\parallel}\;[1 + (g \lambda/\omega_0)^2]\;,\\
\label{eq:effect-of-g-oncouplings} \\ \nonumber
\;\;\;J_{\perp}\;&\rightarrow&\;J^\prime_{\perp} = J_{\perp}\;[1 -
(g \lambda/\omega_0)^2]\;.
\end{eqnarray}

The second effect is the appearance of a term in the Hamiltonian
that breaks the symmetry between the two charge states, favoring
the doubly occupied one:
\begin{equation}
H_{\textsc{akm}}\;\rightarrow\;H^\prime_{\textsc{akm}} -
\frac{g\lambda}{\omega_0}\;J_{\parallel}\tau^z_d\;,
\label{eq:effect-of-g-on-HAKM}
\end{equation}
where $H^\prime_{\textsc{akm}}$ denotes the AKM Hamiltonian with
the new couplings. These effects can be understood quite easily.
In the state with charge $Q_d = 2$ the molecule elongates.
 This
increases its coupling to the electrodes and its energy is lowered
with respect to that of the empty state.  The tunnelling barrier
between them also increases which translates into a reduction of
$J_{\perp}$.

We can use the modified
couplings (\ref{eq:effect-of-g-oncouplings}) to compute the $g$-dependent
Kondo temperature. The result is,
\begin{equation}
{T^\prime_{\textsc{akm}} \over T_{\textsc{akm}}} \sim  \exp\left[
 {\pi g^2 \omega_0
\over \Delta} \left({\lambda \over \omega_0}\right)^4\;
\left(\left(\lambda \over \omega_0\right)^2 - 1\right)\right]\;.
\label{eq:newtakm}
\end{equation}
In the strong coupling regime  the Kondo temperature increases or
 decreases with $g$, depending on the ratio $\lambda/\omega_0$.

The consequences of a finite $g$ on the conductance are twofold.
First, the peak in $G(V_d)$ no longer occurs at $V_d =0$ but it
shifts to $V^\star_d \sim g\;J_{\parallel}\;\lambda/\omega_0$, the
value for which the asymmetry generated by the second term in
Eq.~(\ref{eq:effect-of-g-on-HAKM}) is compensated. Secondly, as in
the previous case, the singly occupied states are more strongly
coupled to the doubly occupied state leading to asymmetry in the
shape of $G(V_d)$. This time, the asymmetry is
 much more pronounced than for weak coupling because the relative weights
of the components of with  $Q_d = 0$ and $Q_d=2$ in the
wavefunction are more sensitive to $V_d$.
\section{Numerical results}
\label{sec:numerical} In this Section we present the numerical
solution of model (\ref{eq:hamil}) using the numerical
renormalization group method
(NRG)~\cite{Wilson1975,Krishnamurthy1980,Costi1994,Hewson_2002} incorporating
a modification designed to improve the accuracy of the computation
of the spectral density. \cite{Hofstetter2000}

Results for the case $g=0$ were discussed in a previous
paper.~\cite{Cornaglia2004} Here, we focus on the effects of
tunnelling barrier modulation
 on the conductance.

In our numerical calculations we used half the bandwidth of the
conduction electron band $D$ as the unit of energy. The parameters
$U=0.1$, $\Delta = 0.01$, and $\omega_0 = 0.05$ were kept fixed
and we studied the conductance as a function of $g$, $\lambda$ and
$V_d$.

\subsection{The electron hole symmetric case at $\lambda =0$}
\label{sec:numerical-g0}

\begin{figure}[tbp]
\includegraphics[width=8.5cm,clip=true]{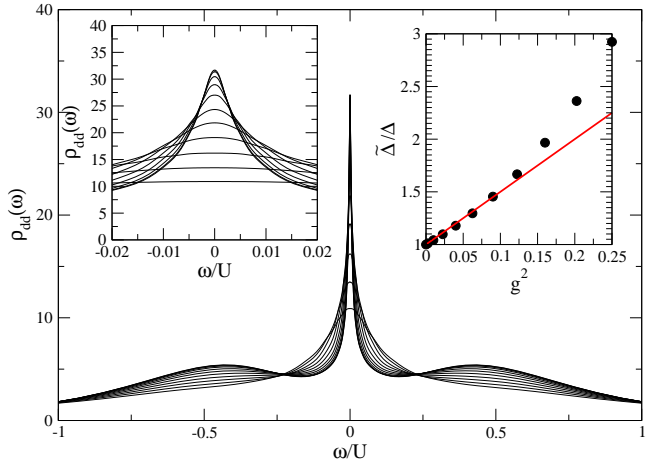}
\caption{Spectral density $\rho_{dd}(\omega)$ at zero temperature
 in the electron-hole symmetric case. The electron-phonon
 coupling $g$ varies between 0 and 0.5 in steps of 0.05
 from top to bottom
in the region of the central peak. Left inset: a zoom of the central part
 of the main figure. Right inset: $g$-dependence of the
effective hybridization parameter $\widetilde{\Delta}$.}
\label{fig:fig2}
\end{figure}

We start by analyzing the electron-hole symmetric case, $\lambda =
0$, $V_d = 0$ [c.f. Eq.~(\ref{eq:Vgate})]. Fig.~\ref{fig:fig2}
shows the  spectral density $\rho_{dd}(\omega)$ at zero
temperature for several values of $g$ between 0 and 0.5.

The top curve corresponds to $g=0$. It exhibits
 a narrow Kondo resonance centered at the Fermi level and two
 Coulomb peaks at $\omega \sim \pm U/2$.
The height of the Kondo resonance is $1/\pi \Delta \sim 32$ in
this case.

With increasing $g$ the hight of the central peak decreases and
its width increases. This evolution is shown in more detail in the
left inset to the figure.

That the height of the central peak depends upon $g$ shows that
$\rho_{dd}(0)$ does not obey Luttinger's theorem
 in the simple form of Eq.~(\ref{eq:Luttinger3}).
An effective hybridization parameter $\widetilde{\Delta}$ may be
obtained from
 the numerical values of $\rho_{dd}(0)$ using Eq.~(\ref{eq:rhodd0}).
 Its dependence on $g$ is represented
 in the right inset to the figure. It is seen that
 $\widetilde{\Delta}$
 increases quadratically with $g$ in the electron-hole symmetric case as was
 anticipated in Section~\ref{sec:Fermi}.

\begin{figure}[tbp]
\includegraphics[width=8.5cm,clip=true]{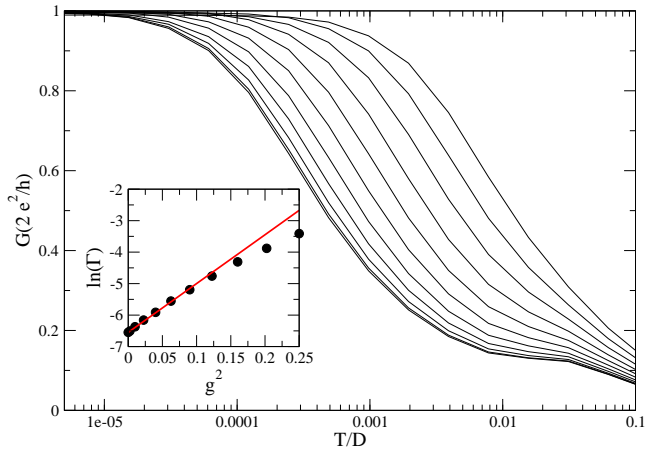}
\caption{Temperature dependence of the conductance in the
electron-hole symmetric case. The electron phonon couplings are as
in Fig.~\ref{fig:fig2}.  The curves correspond to increasing
values of $g$ from left to right. Inset: $g$-dependence of the
width of the central peak of the curves shown in the previous
figure} \label{fig:fig3}
\end{figure}

The figure shows that the width $\Gamma$ of the Kondo resonance
increases monotonically with $g$. The dependence of $\Gamma$ on
$g$ is represented in the inset to Fig.~\ref{fig:fig3} that shows
that $\log \Gamma$ varies quadratically with $g$ for small $g$.
This behavior can be understood from
Eq.~(\ref{eq:Jsmall-lambda-g}), that predicts a quadratic increase
of the Kondo coupling $J$ for small $g$ and $\lambda=0$, recalling
that  $\Gamma \sim T_K$ and that the latter varies exponentially
with $J^{-1}$.

Finally, we observe that the Coulomb peaks broaden with increasing
$g$ until they eventually disappear when the system enters the
mixed-valence regime for large $g$.

The temperature dependence of the conductance at $V_d=0$ is shown
in Fig.~\ref{fig:fig3} for the same values of $g$ as in
Fig.~\ref{fig:fig2}. $G(g,T)$ reaches its maximum value $G_0$ at
$T=0$ for all $g$ in agreement with Eq.~(\ref{eq:rhoddtilde0}).
This confirms the result of Fermi-liquid theory that tunnelling
barrier modulation does not lead to a renormalization of  the
value of the conductance at resonance despite that the spectral
density at the Fermi level is modified.

The conductance at resonance decreases with temperature. The
temperature scale for this process is the Kondo temperature. The
observed shift of the curves to higher temperatures with
increasing $g$ reflects the evolution of the latter.

\subsection{The general case, $\lambda$ and $g \ne 0$}

We pointed out in Section \ref{sec:model} that when both $\lambda$
and $g$ are non zero the system looses electron-hole symmetry.
This is at the origin of several observable features in the
frequency dependence of the molecule's spectral density and in the
dependence of the conductance on gate voltage. All the results
described below were obtained with $g=0.2$.

\begin{figure}[tbp]
\includegraphics[width=8.5cm,clip=true]{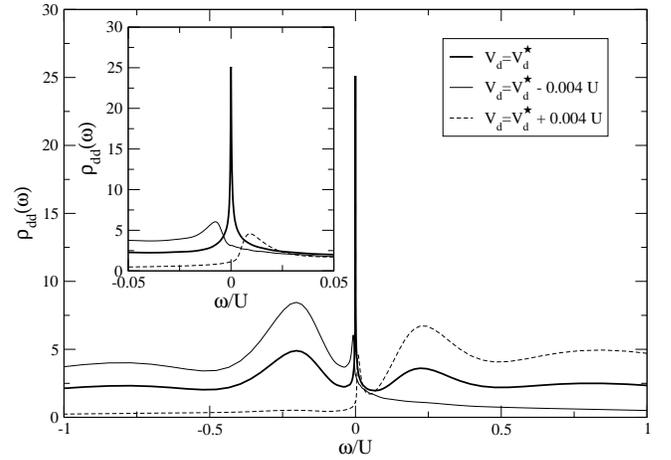}
\caption{Spectral density $\rho_{dd}(\omega)$ at $T=0$ for $\lambda=-0.06$
 and $g=0.2$. Thick line: $V_d/U = V_d^\star/U = 0.159$, $n_d=1$.
Dashed line: $V_d/U = 0.163$, $n_d < 1$. Thin line: $V_d/U = 0.155$,
 $n_d > 1$.
Inset: a zoom of the central peak showing the evolution of
 the Kondo peak with varying $V_d$. }
\label{fig:densesp}
\end{figure}

Figure~\ref{fig:densesp} shows the $d$-electron spectral density
at zero temperature in the strong coupling regime,
$U_{\textrm{eff}} < 0$. We display results for three values of
$V_d$:  $V_d^\star$, chosen so that $n_d = 1$, and $V_d =
V^\star_d \pm \delta V_d$.

The thick line represents $\rho_{dd}(\omega)$ for
 $V_d=V_d^\star$.
We observe a very narrow peak at the
 Fermi level due to the anisotropic Kondo effect described in
Section~\ref{sec:strong-coupling}. We also observe sideband peaks
that are associated to excitation of the higher
 levels of the isolated  molecule at
 $\omega \sim \pm U_{\textrm{eff}}$ [c.f. Eq.~\ref{eq:enes}].
It is seen that these are not positioned  symmetrically with
respect to the central peak and that their widths and heights are
different. As discussed above, the origin of the asymmetry is that
the states with $Q_d = 0$ and $Q_d = 2$ are coupled with different
strength to the states with $Q_d = 1$.  The matrix element for the
transition $|2,0 \rangle \rightarrow |1\sigma,0 \rangle $ is
larger than that for the transition $|0,0 \rangle \rightarrow
|1\sigma,0 \rangle $. 

The spectral densities for $V_d = V^\star_d \pm \delta V_d$ are
represented in the figure by dashed and solid thin lines,
respectively. For our choice of parameters, $|\delta V_d| \lesssim
T_{\textsc{akm}}$. Therefore, the Kondo resonance continues to
exist but it shifts and its amplitude decreases as shown in the
inset to Fig.~\ref{fig:densesp} which is a detailed view of the
central part of the main plot.

When $V_d > V_d^\star$ the empty orbital has a larger weight in
the ground state wavefunction than the doubly occupied orbital.
Then, the sideband at $\omega > 0$ is enhanced and that at $\omega
> 0$ is suppressed. The opposite occurs for $V_d <
V_d^\star$.

For  $|\delta V_d| \gtrsim T_{\textsc{akm}}$ (not shown), the
molecule becomes fully charge-polarized~\cite{Cornaglia2004} and
the Kondo peak is completely suppressed.

We have also computed the spectral density in the weak coupling
regime, $U_{\textrm{eff}}> 0$ (not shown). For moderate values of
$g$, $V_d^\star\sim 0$  and the effects of the
 asymmetry are much less pronounced than for the strong coupling case just discussed.
\begin{figure}[tbp]
\includegraphics[width=8.5cm,clip=true]{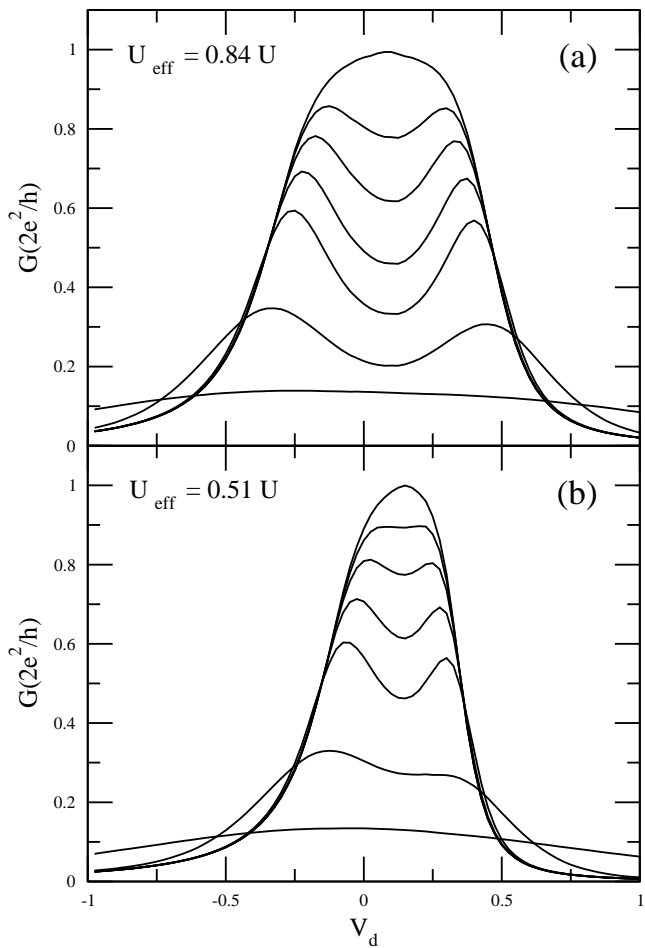}
\caption{Conductance as a function of $V_d$
 and temperature for weak coupling, $U_{\textrm{eff}} > 0$. Panel (a):
 $\lambda = 0.02$. Panel (b): $\lambda = 0.035$. Temperatures are:
 $T=0, 7.3\times 10^{-4}, 1.5\times 10^{-3},3\times 10^{-3},
 6\times 10^{-3}, 0.023$ and 0.094 from top to bottom. In all cases $g=0.2$. }
\label{fig:GVpos}
\end{figure}

We now turn to the analysis of the results for the conductance.
Fig. \ref{fig:GVpos} shows $G(V_d)$ for several temperatures and
two values of $\lambda$, both in the regime $U_{\textrm{eff}}> 0$.
We observe a broad conductance peak of width $\sim
U_{\textrm{eff}}$ at zero temperature. There is perfect
conductance at  $V^\star_d \neq 0$ and the peak is asymmetric with
respect to $V_d^\star$.

At finite temperature the Kondo effect is first suppressed at the
center of the peak where the Kondo temperature is the lowest,
leading to a two-peak structure. There are two
 Coulomb blockade peaks whose widths and hights are different
 for the same reasons that make the spectral density asymmetric.

A comparison between Figs.~\ref{fig:GVpos}(a) and
\ref{fig:GVpos}(b) shows that an increase of $\lambda$ leads to an
enhancement of the anisotropy. This is easily understood as we saw
that the latter arises from the simultaneous presence of ELM and
TBM. We also observe a narrowing of the Coulomb blockade valley
that originates from the decrease of $U_{\textrm{eff}}$.

\begin{figure}[tbp]
\includegraphics[width=8.5cm,clip=true]{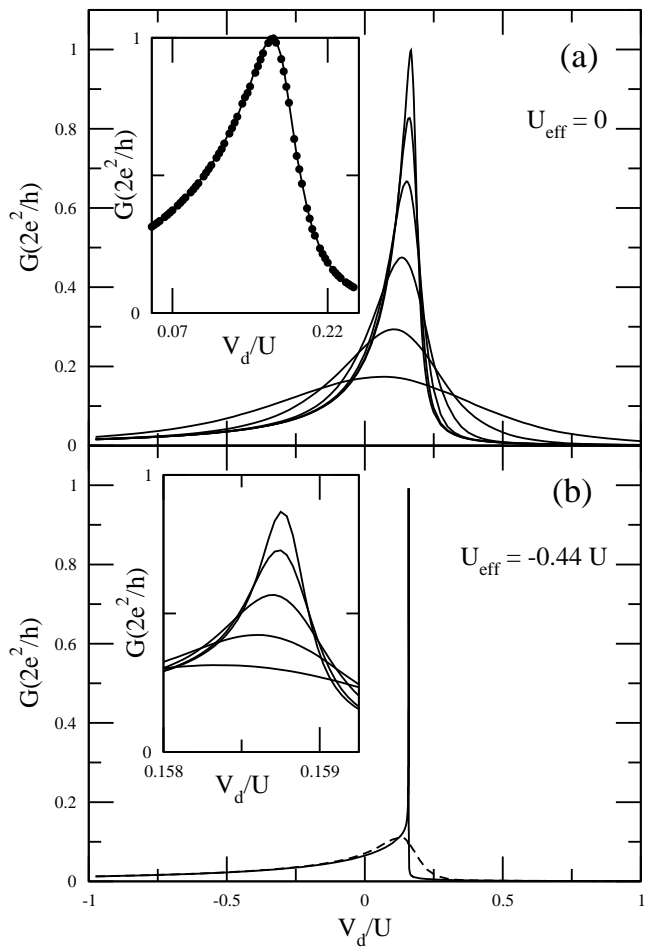}
\caption{Panel (a):  Conductance as a function of $V_d$
 and temperature for $U_{\textrm{eff}} =  0$ ($\lambda = 0.05$).
Temperatures are $T = 0, 3\times 10^{-3}, 6\times 10^{-3},
1.2\times 10^{-2}, 2.4\times 10^{-2}$ and  $4.7\times 10^{-2}$.
Inset:  Check of the
 validity of Luttinger's theorem, Eq.~(\ref{eq:rhoddtilde0}).
Panel (b) Same as (a) for  $U_{\textrm{eff}} <  0$ ($\lambda =
0.06$). Temperatures are $T$=0 and 0.012. Inset. Zoom of the
conductance peak for temperatures $T=1.1\times 10^{-5},
 2.3\times 10^{-5}, 4.6\times 10^{-5}, 9.2\times 10^{-5}$ and
 $1.83\times 10^{-4}$ from top to bottom. For all the curves
 shown in this figure $g=0.2$.}
\label{fig:GVneg}
\end{figure}

A further increase of $\lambda$  takes the molecule
 away from the standard Coulomb blockade regime.
 Fig.~\ref{fig:GVneg} shows results for $U_{\textrm{eff}}=0$. Here,
 the differences between the energies of the four
 charge states of the isolate molecule are
 comparable with their width and no Kondo effect is expected to occur.
 No Coulomb blockade peaks appear upon raising the temperature
 as we argued in Section \ref{sec:strong-coupling}.

The conductance in the strong-coupling case $U_{\textrm{eff}}<0$
is shown in Fig. \ref{fig:GVneg}(b). As in the previous regimes
the zero-temperature conductance is enhanced and reaches
 the quantum of conductance at some value of $V_d$ but, now, the peak is exceedingly narrow.
 This is the regime where the anisotropic Kondo
 effect occurs. The width of the peak in $G(V_d)$ is now
 $T_{\textsc{akm}}$ not
 $|U_{\textrm{eff}}|$.
A very small shift $|\delta V_d| \gtrsim T_{\textsc{akm}}$ is
sufficient to destroy the Kondo resonance. The asymmetry generated
by the tunnelling barrier is in this case extremely pronounced
because the charge on the molecule switches from $n_d =0$ and $n_d
= 2$ within a very small interval of values of $V_d$. Similarly,
the conductance at resonance decreases very rapidly with
increasing temperature as shown in the inset to
Fig.~\ref{fig:GVneg}(b).

We close this Section with a comparison
 between the conductance of the device calculated
from Eq.~(\ref{eq:cond-def}) and that computed from
Eq.~(\ref{eq:rhoddtilde0}), using $n_d(V_d)$ obtained from the
 NRG calculation. The data are shown in the inset to Fig.~\ref{fig:GVneg}(a).
It is seen that the Fermi-liquid relation
 (\ref{eq:rhoddtilde0})  is satisfied within the numerical error.
\section{Conclusions}
\label{sec:conclusions}
 We studied the linear transport properties
of a model molecular transistor in the presence of
electron-electron and electron-phonon interactions analytically
and by numerically solving the model using the Numerical
Renormalization Group method.

Electron-phonon interactions lead to modulation of the positions
of the energy levels of the molecule with respect to the Fermi
level and of the height of the tunnelling barrier between the
molecule and the electrodes. These effects give rise to new
features in the spectral and transport properties of the system.

We found that when there is tunnelling barrier modulation, the
strength of the coupling between the molecule and the leads
depends on the charge of the former. In general this effect breaks
electron-hole symmetry leading to an asymmetric curve of
conductance as a function of gate voltage.

The Kondo effect occurs in the presence of electron-phonon
interactions but its nature is quite different for weak and strong
coupling. In the first regime the ground state of the isolated
molecule is a spin doublet and this leads to a standard Kondo
effect with renormalized parameters. In the second regime the
lowest lying states of the isolated molecule form a charge doublet
and the associated charge fluctuations are described by an
anisotropic Kondo model in a narrow interval of gate voltages. The
two charge states of the doublet are coupled differently to
excited states which gives rise to pronounced asymmetries in the
conductance and in the spectral density.

We established Fermi-liquid relationships for the interacting
model. Tunnelling barrier modulation leads to a non-universal
relation between the spectral density of the molecule and the
electron occupancy. Quite remarkably, the relation between the
zero-temperature conductance and the charge remains unchanged. An
important consequence is that there is perfect transmission in all
regimes when the number of electrons in the molecule is an odd
integer.

\bibliography{refs}

\begin{thebibliography}{27}
\expandafter\ifx\csname natexlab\endcsname\relax\def\natexlab#1{#1}\fi
\expandafter\ifx\csname bibnamefont\endcsname\relax
  \def\bibnamefont#1{#1}\fi
\expandafter\ifx\csname bibfnamefont\endcsname\relax
  \def\bibfnamefont#1{#1}\fi
\expandafter\ifx\csname citenamefont\endcsname\relax
  \def\citenamefont#1{#1}\fi
\expandafter\ifx\csname url\endcsname\relax
  \def\url#1{\texttt{#1}}\fi
\expandafter\ifx\csname urlprefix\endcsname\relax\def\urlprefix{URL }\fi
\providecommand{\bibinfo}[2]{#2}
\providecommand{\eprint}[2][]{\url{#2}}

\bibitem[{\citenamefont{Klein et~al.}(1997)\citenamefont{Klein, Roth, Lim,
  Alivisatos, and McEuen}}]{CB_Qdots}
\bibinfo{author}{\bibfnamefont{D.~L.} \bibnamefont{Klein}},
  \bibinfo{author}{\bibfnamefont{R.}~\bibnamefont{Roth}},
  \bibinfo{author}{\bibfnamefont{A.~K.~L.} \bibnamefont{Lim}},
  \bibinfo{author}{\bibfnamefont{A.~P.} \bibnamefont{Alivisatos}},
  \bibnamefont{and} \bibinfo{author}{\bibfnamefont{P.~L.}
  \bibnamefont{McEuen}}, \bibinfo{journal}{Nature}
  \textbf{\bibinfo{volume}{389}}, \bibinfo{pages}{699} (\bibinfo{year}{1997}).

\bibitem[{\citenamefont{Park et~al.}(2002)\citenamefont{Park, Pasupathy,
  Goldsmith, Chang, Yaish, Petta, Rinkoski, Sethna, Abru{\~n}a, McEuen
  et~al.}}]{JPark_2002}
\bibinfo{author}{\bibfnamefont{J.}~\bibnamefont{Park}},
  \bibinfo{author}{\bibfnamefont{A.~N.} \bibnamefont{Pasupathy}},
  \bibinfo{author}{\bibfnamefont{J.~I.} \bibnamefont{Goldsmith}},
  \bibinfo{author}{\bibfnamefont{C.}~\bibnamefont{Chang}},
  \bibinfo{author}{\bibfnamefont{Y.}~\bibnamefont{Yaish}},
  \bibinfo{author}{\bibfnamefont{J.~R.} \bibnamefont{Petta}},
  \bibinfo{author}{\bibfnamefont{M.}~\bibnamefont{Rinkoski}},
  \bibinfo{author}{\bibfnamefont{J.~P.} \bibnamefont{Sethna}},
  \bibinfo{author}{\bibfnamefont{H.~D.} \bibnamefont{Abru{\~n}a}},
  \bibinfo{author}{\bibfnamefont{P.~L.} \bibnamefont{McEuen}},
  \bibnamefont{et~al.}, \bibinfo{journal}{Nature}
  \textbf{\bibinfo{volume}{417}}, \bibinfo{pages}{722} (\bibinfo{year}{2002}).

\bibitem[{\citenamefont{Kubatkin et~al.}(2003)\citenamefont{Kubatkin, Danilov,
  Hjort, Cornil, Br{\'e}das, Stuhr-Hansen, Hedeg{\aa}rd, and
  Bj{\o}rnholm}}]{Kubatkin_2003}
\bibinfo{author}{\bibfnamefont{S.}~\bibnamefont{Kubatkin}},
  \bibinfo{author}{\bibfnamefont{A.}~\bibnamefont{Danilov}},
  \bibinfo{author}{\bibfnamefont{M.}~\bibnamefont{Hjort}},
  \bibinfo{author}{\bibfnamefont{J.}~\bibnamefont{Cornil}},
  \bibinfo{author}{\bibfnamefont{J.-L.} \bibnamefont{Br{\'e}das}},
  \bibinfo{author}{\bibfnamefont{N.}~\bibnamefont{Stuhr-Hansen}},
  \bibinfo{author}{\bibfnamefont{P.}~\bibnamefont{Hedeg{\aa}rd}},
  \bibnamefont{and}
  \bibinfo{author}{\bibfnamefont{T.}~\bibnamefont{Bj{\o}rnholm}},
  \bibinfo{journal}{Nature} \textbf{\bibinfo{volume}{425}},
  \bibinfo{pages}{698} (\bibinfo{year}{2003}).

\bibitem[{\citenamefont{Goldhaber-Gordon
  et~al.}(1998)\citenamefont{Goldhaber-Gordon, Shtrikman, Mahalu,
  Abusch-Magder, Meirav, and Kastner}}]{Gordon1998}
\bibinfo{author}{\bibfnamefont{D.}~\bibnamefont{Goldhaber-Gordon}},
  \bibinfo{author}{\bibfnamefont{H.}~\bibnamefont{Shtrikman}},
  \bibinfo{author}{\bibfnamefont{D.}~\bibnamefont{Mahalu}},
  \bibinfo{author}{\bibfnamefont{D.}~\bibnamefont{Abusch-Magder}},
  \bibinfo{author}{\bibfnamefont{U.}~\bibnamefont{Meirav}}, \bibnamefont{and}
  \bibinfo{author}{\bibfnamefont{M.~A.} \bibnamefont{Kastner}},
  \bibinfo{journal}{Nature} \textbf{\bibinfo{volume}{391}},
  \bibinfo{pages}{156} (\bibinfo{year}{1998}).

\bibitem[{\citenamefont{Liang et~al.}(2002)\citenamefont{Liang, Shores,
  Bockrath, Long, and Park}}]{WLiang_2002}
\bibinfo{author}{\bibfnamefont{W.}~\bibnamefont{Liang}},
  \bibinfo{author}{\bibfnamefont{M.~P.} \bibnamefont{Shores}},
  \bibinfo{author}{\bibfnamefont{M.}~\bibnamefont{Bockrath}},
  \bibinfo{author}{\bibfnamefont{J.~R.} \bibnamefont{Long}}, \bibnamefont{and}
  \bibinfo{author}{\bibfnamefont{H.}~\bibnamefont{Park}},
  \bibinfo{journal}{Nature} \textbf{\bibinfo{volume}{417}},
  \bibinfo{pages}{725} (\bibinfo{year}{2002}).

\bibitem[{\citenamefont{Yu and Natelson}(2004)}]{Yu2004}
\bibinfo{author}{\bibfnamefont{L.~H.} \bibnamefont{Yu}} \bibnamefont{and}
  \bibinfo{author}{\bibfnamefont{D.}~\bibnamefont{Natelson}},
  \bibinfo{journal}{Nano Letters} \textbf{\bibinfo{volume}{4(1)}},
  \bibinfo{pages}{79} (\bibinfo{year}{2004}).

\bibitem[{\citenamefont{Yu et~al.}()\citenamefont{Yu, Keane, Ciszek, Cheng,
  Stewart, Tour, and Natelson}}]{Yu2004b}
\bibinfo{author}{\bibfnamefont{L.~H.} \bibnamefont{Yu}},
  \bibinfo{author}{\bibfnamefont{Z.~K.} \bibnamefont{Keane}},
  \bibinfo{author}{\bibfnamefont{J.~W.} \bibnamefont{Ciszek}},
  \bibinfo{author}{\bibfnamefont{L.}~\bibnamefont{Cheng}},
  \bibinfo{author}{\bibfnamefont{M.~P.} \bibnamefont{Stewart}},
  \bibinfo{author}{\bibfnamefont{J.~M.} \bibnamefont{Tour}}, \bibnamefont{and}
  \bibinfo{author}{\bibfnamefont{D.}~\bibnamefont{Natelson}},
  \bibinfo{note}{cond-nat/0408052}.

\bibitem[{\citenamefont{Ho}(2002)}]{WHoXX}
\bibinfo{author}{\bibfnamefont{W.}~\bibnamefont{Ho}}, \bibinfo{journal}{J.
  Chem. Phys.} \textbf{\bibinfo{volume}{55}}, \bibinfo{pages}{11033}
  (\bibinfo{year}{2002}).

\bibitem[{\citenamefont{Park et~al.}(2000)\citenamefont{Park, Park, Lim,
  Anderson, Alivisatos, and McEuen}}]{HPark_2000}
\bibinfo{author}{\bibfnamefont{H.}~\bibnamefont{Park}},
  \bibinfo{author}{\bibfnamefont{J.}~\bibnamefont{Park}},
  \bibinfo{author}{\bibfnamefont{A.~K.~L.} \bibnamefont{Lim}},
  \bibinfo{author}{\bibfnamefont{E.~H.} \bibnamefont{Anderson}},
  \bibinfo{author}{\bibfnamefont{A.~P.} \bibnamefont{Alivisatos}},
  \bibnamefont{and} \bibinfo{author}{\bibfnamefont{P.~L.}
  \bibnamefont{McEuen}}, \bibinfo{journal}{Nature}
  \textbf{\bibinfo{volume}{407}}, \bibinfo{pages}{57} (\bibinfo{year}{2000}).

\bibitem[{\citenamefont{Weig et~al.}(2004)\citenamefont{Weig, Blick, Brandes,
  Kirschbaum, Wegscheider, Bichler, and Kotthaus}}]{Weig2004}
\bibinfo{author}{\bibfnamefont{E.~M.} \bibnamefont{Weig}},
  \bibinfo{author}{\bibfnamefont{R.~H.} \bibnamefont{Blick}},
  \bibinfo{author}{\bibfnamefont{T.}~\bibnamefont{Brandes}},
  \bibinfo{author}{\bibfnamefont{J.}~\bibnamefont{Kirschbaum}},
  \bibinfo{author}{\bibfnamefont{W.}~\bibnamefont{Wegscheider}},
  \bibinfo{author}{\bibfnamefont{M.}~\bibnamefont{Bichler}}, \bibnamefont{and}
  \bibinfo{author}{\bibfnamefont{J.~P.} \bibnamefont{Kotthaus}},
  \bibinfo{journal}{Phys. Rev. Lett.} \textbf{\bibinfo{volume}{92}},
  \bibinfo{pages}{046804} (\bibinfo{year}{2004}).

\bibitem[{\citenamefont{Mitra et~al.}(2004)\citenamefont{Mitra, Aleiner, and
  Millis}}]{Mitra_2004}
\bibinfo{author}{\bibfnamefont{A.}~\bibnamefont{Mitra}},
  \bibinfo{author}{\bibfnamefont{I.}~\bibnamefont{Aleiner}}, \bibnamefont{and}
  \bibinfo{author}{\bibfnamefont{A.}~\bibnamefont{Millis}},
  \bibinfo{journal}{Phys. Rev. B} \textbf{\bibinfo{volume}{69}},
  \bibinfo{pages}{245302} (\bibinfo{year}{2004}).

\bibitem[{\citenamefont{M.-T.Kuo and Chang}(2002)}]{Kuo_2002}
\bibinfo{author}{\bibfnamefont{D.}~\bibnamefont{M.-T.Kuo}} \bibnamefont{and}
  \bibinfo{author}{\bibfnamefont{Y.~C.} \bibnamefont{Chang}},
  \bibinfo{journal}{Phys. Rev. B} \textbf{\bibinfo{volume}{66}}
  (\bibinfo{year}{2002}).

\bibitem[{\citenamefont{Flensberg}(2003)}]{Flensberg_2004}
\bibinfo{author}{\bibfnamefont{K.}~\bibnamefont{Flensberg}},
  \bibinfo{journal}{Phys. Rev. B} \textbf{\bibinfo{volume}{68}},
  \bibinfo{pages}{205323} (\bibinfo{year}{2003}).

\bibitem[{\citenamefont{Cornaglia et~al.}(2004)\citenamefont{Cornaglia, Ness,
  and Grempel}}]{Cornaglia2004}
\bibinfo{author}{\bibfnamefont{P.~S.} \bibnamefont{Cornaglia}},
  \bibinfo{author}{\bibfnamefont{H.}~\bibnamefont{Ness}}, \bibnamefont{and}
  \bibinfo{author}{\bibfnamefont{D.~R.} \bibnamefont{Grempel}},
  \bibinfo{journal}{To appear in Phys. Rev. Lett.}  (\bibinfo{year}{2004}).

\bibitem[{\citenamefont{Arrachea et~al.}(2003)\citenamefont{Arrachea, Aligia,
  and Santoro}}]{Arrachea2003}
\bibinfo{author}{\bibfnamefont{L.}~\bibnamefont{Arrachea}},
  \bibinfo{author}{\bibfnamefont{A.~A.} \bibnamefont{Aligia}},
  \bibnamefont{and} \bibinfo{author}{\bibfnamefont{G.~E.}
  \bibnamefont{Santoro}}, \bibinfo{journal}{Phys. Rev. B}
  \textbf{\bibinfo{volume}{67}}, \bibinfo{pages}{134307}
  (\bibinfo{year}{2003}).

\bibitem[{\citenamefont{Meir and Wingreen}(1992)}]{Meir1992}
\bibinfo{author}{\bibfnamefont{Y.}~\bibnamefont{Meir}} \bibnamefont{and}
  \bibinfo{author}{\bibfnamefont{N.~S.} \bibnamefont{Wingreen}},
  \bibinfo{journal}{Phys. Rev. Lett.} \textbf{\bibinfo{volume}{68}},
  \bibinfo{pages}{2512} (\bibinfo{year}{1992}).

\bibitem[{\citenamefont{Jauho et~al.}(1994)\citenamefont{Jauho, Wingreen, and
  Meir}}]{Jauho1994}
\bibinfo{author}{\bibfnamefont{A.-P.} \bibnamefont{Jauho}},
  \bibinfo{author}{\bibfnamefont{N.~S.} \bibnamefont{Wingreen}},
  \bibnamefont{and} \bibinfo{author}{\bibfnamefont{Y.}~\bibnamefont{Meir}},
  \bibinfo{journal}{Phys. Rev. B} \textbf{\bibinfo{volume}{50}},
  \bibinfo{pages}{5528} (\bibinfo{year}{1994}).

\bibitem[{\citenamefont{Abrikosov et~al.}(1963)\citenamefont{Abrikosov, Gorkov,
  and Dzyaloshinski}}]{Abrikosov-book}
\bibinfo{author}{\bibfnamefont{A.~A.} \bibnamefont{Abrikosov}},
  \bibinfo{author}{\bibfnamefont{L.~P.} \bibnamefont{Gorkov}},
  \bibnamefont{and} \bibinfo{author}{\bibfnamefont{I.~E.}
  \bibnamefont{Dzyaloshinski}}, \emph{\bibinfo{title}{Methods of Quantum Field
  Theory in Statistical Physics}} (\bibinfo{publisher}{Prentice-Hall},
  \bibinfo{address}{Englewood-Cliffs, New Jersey}, \bibinfo{year}{1963}).

\bibitem[{\citenamefont{Hewson}(1997)}]{Hewson-book}
\bibinfo{author}{\bibfnamefont{A.~C.} \bibnamefont{Hewson}},
  \emph{\bibinfo{title}{The Kondo Problem to Heavy Fermions}}
  (\bibinfo{publisher}{Cambridge University Press}, \bibinfo{year}{1997}).

\bibitem[{\citenamefont{Costi and Kieffer}(1996)}]{Costi1996}
\bibinfo{author}{\bibfnamefont{T.~A.} \bibnamefont{Costi}} \bibnamefont{and}
  \bibinfo{author}{\bibfnamefont{C.}~\bibnamefont{Kieffer}},
  \bibinfo{journal}{Phys. Rev. Lett.} \textbf{\bibinfo{volume}{76}},
  \bibinfo{pages}{1683} (\bibinfo{year}{1996}).

\bibitem[{\citenamefont{Costi}(1998)}]{Costi1998}
\bibinfo{author}{\bibfnamefont{T.~A.} \bibnamefont{Costi}},
  \bibinfo{journal}{Phys. Rev. Lett.} \textbf{\bibinfo{volume}{80}},
  \bibinfo{pages}{1038} (\bibinfo{year}{1998}).

\bibitem[{\citenamefont{Wilson}(1975)}]{Wilson1975}
\bibinfo{author}{\bibfnamefont{K.~G.} \bibnamefont{Wilson}},
  \bibinfo{journal}{Rev. Mod. Phys.} \textbf{\bibinfo{volume}{47}},
  \bibinfo{pages}{773} (\bibinfo{year}{1975}).

\bibitem[{\citenamefont{Krishna-murthy
  et~al.}(1980)\citenamefont{Krishna-murthy, Wilkins, and
  Wilson}}]{Krishnamurthy1980}
\bibinfo{author}{\bibfnamefont{H.~R.} \bibnamefont{Krishna-murthy}},
  \bibinfo{author}{\bibfnamefont{J.}~\bibnamefont{Wilkins}}, \bibnamefont{and}
  \bibinfo{author}{\bibfnamefont{K.~G.} \bibnamefont{Wilson}},
  \bibinfo{journal}{Phys. Rev. B} \textbf{\bibinfo{volume}{21}},
  \bibinfo{pages}{1003} (\bibinfo{year}{1980}).

\bibitem[{\citenamefont{Costi et~al.}(1994)\citenamefont{Costi, Hewson, and
  Zlati{\'c}}}]{Costi1994}
\bibinfo{author}{\bibfnamefont{T.~A.} \bibnamefont{Costi}},
  \bibinfo{author}{\bibfnamefont{A.~C.} \bibnamefont{Hewson}},
  \bibnamefont{and}
  \bibinfo{author}{\bibfnamefont{V.}~\bibnamefont{Zlati{\'c}}},
  \bibinfo{journal}{J. Phys.:Condens. Matter} \textbf{\bibinfo{volume}{6}},
  \bibinfo{pages}{2519} (\bibinfo{year}{1994}).

\bibitem[{\citenamefont{Hewson and Meyer}(2002)}]{Hewson_2002}
\bibinfo{author}{\bibfnamefont{A.~C.} \bibnamefont{Hewson}} \bibnamefont{and}
  \bibinfo{author}{\bibfnamefont{D.}~\bibnamefont{Meyer}}, \bibinfo{journal}{J.
  Phys.:Condens. Matter} \textbf{\bibinfo{volume}{14}}, \bibinfo{pages}{427}
  (\bibinfo{year}{2002}).

\bibitem[{\citenamefont{Hofstetter}(2000)}]{Hofstetter2000}
\bibinfo{author}{\bibfnamefont{W.}~\bibnamefont{Hofstetter}},
  \bibinfo{journal}{Phys. Rev. Lett.} \textbf{\bibinfo{volume}{85}},
  \bibinfo{pages}{1508} (\bibinfo{year}{2000}).

\bibitem[{\citenamefont{Stephan et~al.}(1997)\citenamefont{Stephan, Capone,
  Grilli, and Castellani}}]{Stephan1997}
\bibinfo{author}{\bibfnamefont{W.}~\bibnamefont{Stephan}},
  \bibinfo{author}{\bibfnamefont{M.}~\bibnamefont{Capone}},
  \bibinfo{author}{\bibfnamefont{M.}~\bibnamefont{Grilli}}, \bibnamefont{and}
  \bibinfo{author}{\bibfnamefont{C.}~\bibnamefont{Castellani}},
  \bibinfo{journal}{Phys. Lett. A} \textbf{\bibinfo{volume}{227 (1-2)}}
  (\bibinfo{year}{1997}).

\end{thebibliography}
\end{document}